\documentclass[12pt,a4paper,final]{iopart}
\newcommand{\be}{\begin{equation}}
\newcommand{\ee}{\end{equation}}
\newcommand{\bea}{\begin{eqnarray}}
\newcommand{\eea}{\end{eqnarray}}
\newcommand{\ba}[1]{\begin{array}{#1}}
\newcommand{\ea}{\end{array}}
\usepackage{amssymb}
\usepackage{mathrsfs}
\usepackage{color}
\usepackage{iopams} 
\usepackage{graphicx}
\usepackage[breaklinks=true,colorlinks=true,linkcolor=blue,urlcolor=blue,citecolor=blue]{hyperref}
\newcommand{\threej}[6]{
\ensuremath{
\left( \!\!
\begin{array}{ccc}
#1 & #2 & #3 \\
#4 & #5 & #6 \\
\end{array}
\!\!\right) 
}}

\begin{document}
\setlength{\topmargin}{0.2in}

\title[nanoscale atom-atom interactions with CQED]{Manipulating nanoscale atom-atom interactions with cavity QED}

\author{Arpita Pal$^1$, Subrata Saha$^1$ and Bimalendu Deb$^{1,2}$}
\address{$^1$Department of Materials Science, Indian Association for the Cultivation of Science, Jadavpur, Kolkata 700032, INDIA.}
\address{$^2$Raman Centre for Atomic, Molecular and Optical Sciences, Indian Association for the Cultivation of Science, Jadavpur, Kolkata 700032, INDIA.}

\begin{abstract}
We theoretically explore manipulation of interactions between excited and ground state atoms at nanoscale separations by cavity quantum electrodynamics (CQED). We develop an adiabatic molecular dressed state formalism and show that it is possible to generate Fano-Feshbach resonances between ground and long-lived excited-state atoms inside a cavity. The resonances are shown to arise due to non-adiabatic coupling near a pseudo-crossing between the dressed state potentials. We illustrate our results with a model study using fermionic $^{171}$Yb atoms in a two-modal cavity. Our study is important for manipulation of interatomic interactions at low energy by cavity field.
\end{abstract}

\pacs{34.10.+x, 42.50.Pq, 33.80.-b}

\maketitle

\section{Introduction}
Cavity quantum electrodynamics (CQED)\cite{berman,haroche} deals primarily with atom-photon interactions in a fully quantum mechanical way, without much recourse to atom-atom interactions. Since Purcell's celebrated work \cite{purcell} about 70 year ago, showing modification of spontaneous emission via tailoring of vacuum electrodynamics modes with a cavity, CQED has been traditionally developed as an area of fundamental research in light-matter interactions. A paradigmatic exactly solvable model  in this field is Jaynes-Cummings model \cite{jaynes} that describes interaction  of a  single two-level atom (TLA)  with a single-mode cavity field. A variety of models \cite{tavis} of an ensemble of non-interacting atoms collectively interacting with a single-mode quantized field has been used over the years for studying collective effects in CQED. With the advent of Bose-Einstein condensates (BEC) \cite{bec} of atomic gases about twenty years back, collective QED effects  with ultracold atoms inside a cavity have 
become important \cite{collective:Esslinger,collective}. In such collective systems, interatomic or molecular interactions are generally ignored because the  interatomic separation is  usually quite large compared to typical size for molecular interactions. However, cavity photon-mediated long-range correlations or interactions \cite{ddi,Zheng:2000,Petrosyan:2003,Esslinger:Science:2012,Ritsch:rmp:2013,Blatt:prl:2013,reimannprl2015} between atoms have attracted a lot of research interests in recent times. Of late, long-range resonant dipole-dipole interactions (RDDI) mediated by real or virtual photons in a cavity or waveguide have become important \cite{kobayashipra1995, agarwalpra1998, Lawandyjmo1994, huangpra2012, johnnjp2013, kurizkipra2014}. In this work, we do not consider such long range interactions, rather we primarily focus on relatively short-range interactions.

One of the principal aims of CQED studies is to attain  extraordinary control over the atomic and photonic states  for fundamental and quantum information studies\cite{Haroche:prl:1999,Kimble:prl:2007,Rempe:prl:2011-15}. This is accomplished by controlling the spontaneous emission processes of atoms by cavity. Since it is generally difficult to attain such control over interatomic or molecular interactions, molecular physics has not so far found much inroads into CQED. Nevertheless, several theoretical  \cite{debprl99,nussenzveig:2001,ritsch:2005,pinske:2007,zhou:2008} and experimental  \cite{Rempe:PRL:2000,Haroche:2001-2,Ritsch:rmp:2013,reimannprl2015,Srivathsan:prl:2013,Vuletic:Nature:2013} works have been performed towards this direction.

Here we carry out a model study to show that it is possible to manipulate atom-atom interactions and  excited-state potentials at nanoscale separations using CQED. As a model system, we consider interactions between two colliding V-type atoms inside a two-mode cavity. We work in the basis of adiabatic molecule-cavity dressed states in the center-of-mass molecular frame of reference. We then investigate into the cavity-modified interactions between  one ground- and  the other excited-state atoms. We study the non-adiabatic effects near a pseudo-crossing between dressed-state potentials at a nanoscale separation. When the upper adiabatic dressed-state potential of a pseudo-crossing is a binding potential, the modification of interactions between the atoms is shown to occur due to Fano effect \cite{fanopr1961}. The non-adiabatic coupling between a  bound-state supported by the upper potential and the continuum of states in the lower potential leads to Fano resonances. In the asymptotic limit, the 
two potential curves correspond to two separated atoms of which one is in the excited state. We consider those kind of atoms which have either very long life time (typically in microsecond regime) or meta-stable excited states such as alkaline-earth metal atoms or other two valance electron systems such as ytterbium. In recent times,  cold collision in meta-stable excited states has become important \cite{takahashi2016}.

In this paper we are interested in nanoscale CQED effects on purely long range (PLR) interaction between  two atoms. PLR potentials arise due to combined effects of a number of diatomic interactions such as resonant dipole-dipole, molecular spin-orbit, hyperfine and quadrupole interactions at separations beyond the chemically active zone of overlapping charge clouds of the two atoms. These potentials have prominent effects, such as binding between two atoms forming exotic bound states typically at a few nanometer separations \cite{enomotoprl2008}. Many atomic species such as K, Na, He, Cs, Rb have well defined PLR states \cite{plr}. For numerical illustration of our theory, we choose relatively simple two valance electron fermionic $^{171}$Yb atoms which have nuclear spin 1/2.  The atomic ground-state  of $^{171}$Yb is purely electronic spin-singlet. $^{171}$Yb$_2$  has  excited  PLR states \cite{enomotoprl2008} that are accessible via $^1S_0$ -$^3P_1$ intercombination transition. In this system, PLR states 
appear 
due to an interplay between reson dipole-dipole interaction (RDDI) and hyperfine  interaction. Furthermore, $^{171}$Yb is useful for cavity QED experiments. In a recent experiment using two-mode cavity QED set up, Eto {\it et al.} \cite{ueda:prl:2011}  have  demonstrated that the  nuclear spin of $^{171}$Yb can significantly influence  cavity-enhanced fluorescence. Recent progress in optical control of atom-atom interactions at nanometer scale \cite{enomotoprl2008,pichlerprl1978,pra:2010:goel} using narrowline intercombination free-bound transition motivates us to explore theoretically CQED effects on nanoscale interactions between cold atoms.

For experimental realization of our proposal, we envisage a situation where a dense cloud of cold atoms can be loaded into cavity field standing wave such that the pairs of atoms can be trapped in the antinode of the standing wave. In fact in recent times, there has been tremendous progress in cavity cooling of atoms \cite{cavitycool}. So it is expected that in near future a large number of atoms can be cooled and trapped by a cavity field. This will be like a cavity generated optical lattice \cite{cavitylattice,Ritsch:rmp:2013}. In case of a cavity lattice with high filling factor it is likely that the pairs of atoms can be localized in an antinode of the standing wave. Alternatively, for initial preparation of the system with a pair of atoms in close proximity, one can imagine that a Mott-insulator \cite{mot:2002,mot:2016} with doubly occupied sites can be loaded into an empty cavity. Then after excitations of cavity fields one can extinguish the Mott-insulator lattice.

The paper is organized in the following way. In Sec.2. we present our model consisting of a pair of V-type three level atoms with long-lived or metastable excited states interacting with a two-mode quantized cavity field. We develop an adiabatic dressed state formalism using diatom-field coupled Bell-type basis in Sec.3. We apply this formalism to solve our model and analyze the non-adiabatic effects near a pseudo-crossing. In particular, we calculate the nonadiabatic effect-induced Fano-Feshbach resonances in  scattering between ground- and  long-lived excited-state atoms. We discuss our numerical results in Sec.4. In Sec.5, we conclude this paper.

\begin{figure}
 \includegraphics[width=0.9\linewidth]{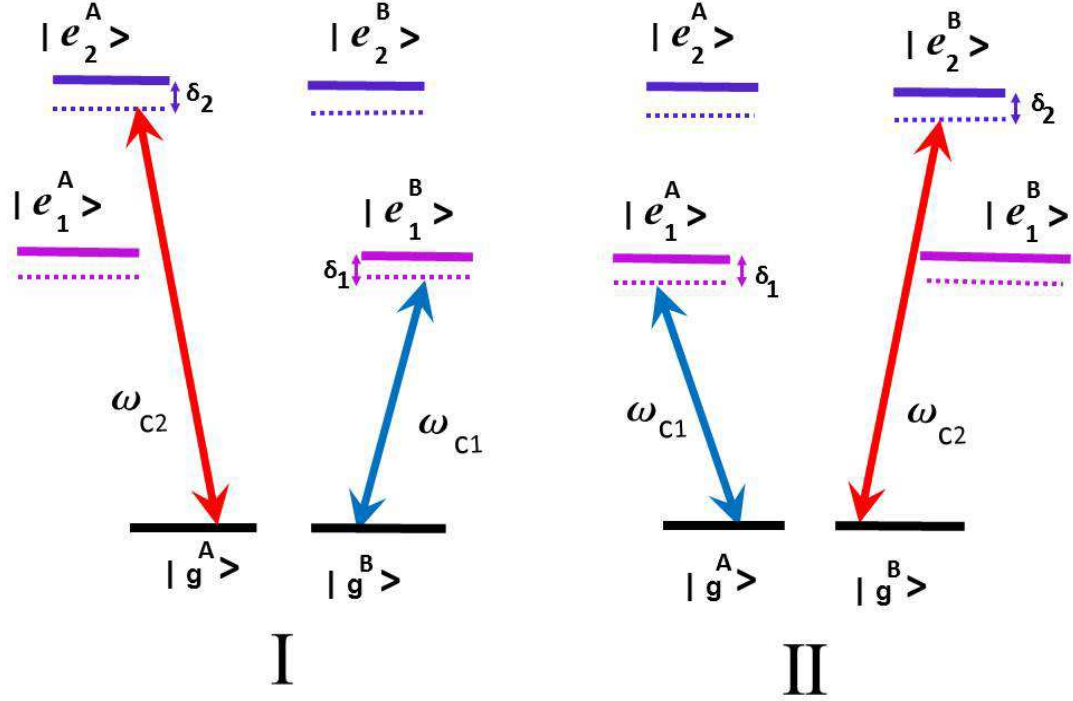}
 \caption{A schematic diagram of two atoms A and B in V-type configuration inside a two-mode cavity. $|e^i_1\rangle$ and $|e^i_2\rangle$ are the two sublevels of the excited state and the ground ($^1$S$_0$) state of $i$-th ($i = A, B$) atom. Blue and red arrows indicate cavity photons of the mode frequencies $\omega_{c1}$ and $\omega_{c2}$, respectively. Here I and II show two possible pathways for two photons to interact with both atoms.}
 \label{fig1}
\end{figure}
\section{The Model Hamiltonian}
We consider a pair of interacting V-type three-level atoms placed at about nanoscale separation. To begin with, we keep our discussion most general. The hamiltonian $\hat {H}$ contains two parts:  $\hat{H} = \hat{H}_{{\rm k}} + \hat{H}_{{\rm ad}}$ where $\hat{H}_{{\rm k}}$ and $\hat{H}_{{\rm ad}}$ represent the kinetic and adiabatic Hamiltonian. The kinetic part can be written as 
\bea
\hat{H}_{\rm k} = \sum_{\mu, \nu} \left [ -\frac{\hbar^2}{2\mu_m}\frac{d^2}{dR^2} + 
\frac{\hbar^2}{ 2 \mu R^2} \left \{ \left \langle \, \left (  {\mathbf J}_{\mu \nu}  -  
{\mathbf J}_{\mu \nu }^{{\rm elec}} \right )^2  \, \right \rangle 
\right \} \right ]\mid \mu, \nu \rangle \langle \mu, \nu \mid 
\label{eq1}
\eea 
where $\mu_m = \frac{m_A m_B}{m_A + m_B}$ is the reduced mass of the two atoms, $R = |\vec{R_A} - \vec{R_B}|$ is the separation between the atoms, ${\mathbf J}_{\mu \nu}$ is the total molecular angular momentum including electronic orbital (L) and electronic spin (S), nuclear spin (I) and the rotational ($\ell$) motion of the inter-nuclear axis, ${\mathbf J}_{\mu \nu }^{{\rm elec}}$ denotes molecular electronic angular momentum. The symbol $\langle \cdots \rangle $ implies averaging over the angular states in the center-of-mass (COM) frame. The subscripts $\mu \nu$ indicate that the  angular momenta correspond to the molecular states which asymptotically correspond to  $\mid \mu, \nu \rangle$. Now, the adiabatic hamiltonian $\hat{H}_{ad}$ (it is adiabatic in the sense that relative and the COM motion of the two atoms are not taken into account) can be written as
\bea
\hat{H}_{ad} =  \hat{H}_{0}+ \hat{H}_{\rm af} + \hat{H}_{\rm RDDI} + \hat{H}_{\rm hf}  + \hat{V}_{\rm LR} 
\label{eq2}
\eea 
where
\bea
\hat{H}_{0} = \sum_{i'=A,B} \sum^{2}_{j'=1} \hbar (\Omega^{i'}_{j'} \mid e^{i'}_{j'}\rangle \langle e^{i'}_{j'} \mid )+ \hbar \sum_{k=1,2} N_k\omega_{c_k} \hat{a}^\dag_k \hat{a}_k 
\label{eq3}
\eea
is the free part with $\hat{a_k}$ being the annihilation operators of the cavity field mode $k$ with $N_k$ photons. Here $\mid e^{i'}_{j'}\rangle$ represents excited atomic state of ${i'}^{th}$ atom in ${j'}^{th}$ level.
\bea
\hat{H}_{\rm af} = \sum_{i'=A,B}\sum^{2}_{j'=1}\sum^2_{k=1} \hbar({\rm{g}}_{i'}\hat{a}_{k}\mid e^{i'}_{j'}\rangle \langle {g}^{i'}\mid \hskip.1cm \delta_{kj'}+{\rm H.c.})
\label{eq6}
\eea
where $\rm{g}_{i'}$ is the atom-field coupling parameter of ${i'}^{\rm th}$ atom. $ \mid g^{i'}\rangle$ represents the ground atomic state of ${i'}^{th}$ atom. 
\bea
\hat{H}_{\rm RDDI} = \frac{1}{R^3}\sum_{j'=1,2; j''=1,2}  C_{j' j''} \mid e^{A}_{j'}, g^{B}\rangle \langle e^{B}_{j''}, g^{A}\mid + {\rm H.c.}
\label{eq4}
\eea
Here $\hat{H}_{\rm RDDI}$ describes the resonant dipole-dipole interaction interaction (RDDI) where $\mid e^{A}_{j'},g^{B}\rangle$ is the product of atomic states of atoms $A$ and $B$ and  $C_{j'j''}$ is the dipole-dipole coupling coefficient. In 1930, London  \cite{londonzfp1930} showed that, at fairly large separation between two homo-nuclear atoms the interatomic potential energy varies as $1/R^3$ (where R is the separation between the atoms) when one atom is in excited state and other is in ground state. This potentials results from RDDI mediated by a single-photon between the excited- and the ground-state atoms. The  is described by the hamiltonian,
where $\hat{H}_{\rm hf}$ is the hyperfine interaction
\bea
\hat{H}_{\rm hf} = \sum_{i'=A,B}\sum^2_{j'=1} a_{\rm hf}( \vec{i_{j'}}.\vec{j_{j'}})\mid e^{i'}_{j'}\rangle \langle e^{i'}_{j'} \mid
\label{eq5}
\eea
where $a_{\rm hf}$ is the atomic hyperfine constant, $\vec{i_{j'}}$ and $\vec{j_{j'}}$ are the nuclear spin and total electronic angular momentum of ${j'}^{th}$ level of atom $i'$. Besides these interactions there are other long range interaction potentials which are important for understanding long-range forces between the two atoms. Those can be expressed as
\bea 
\hat{V}_{\rm LR} &=& V_{gg}(R) \mid g^{A},g^{B}\rangle \langle g^{A},g^{B} \mid + V_{e g}(R) \sum_{j'} \mid e^{A}_{j'},g^{B}\rangle \langle e^{A}_{j'},g^{B} \mid \nonumber \\
&+& V_{g e}(R) \sum_{j'} \mid g^{A},e^{B}_{j'}\rangle \langle g^{A},e^{B}_{j'} \mid + V_{QQ} (R) \sum_{j' j''} \mid e^{A}_{j'},e_{j''}^{B}\rangle \langle e_{j'}^{A},e_{j''}^{B} \mid 
\eea 
where $V_{gg}(R)$, $V_{e g}(R)$ and $V_{g e}(R)$ represent ground and excited state potentials which at long separation behave as $\sim 1/R^6$ as $R \rightarrow \infty$. $V_{QQ}(R)$ is the first order correction to molecular term i.e. the quadrupole interaction.

\section{Adiabatic Dressed-state formalism}
To construct suitable atomic or molecular basis for our system, we choose product representation of atomic states within the framework of Movre-Pichlar model \cite{movre}. At purely long range or at nm separation the interatomic central or electrostatic interaction can be treated as perturbation compared to the resonant dipole dipole interaction. Hence a molecular state can be well described by this model. For two spin-polarized $^{171}$Yb atoms, the ground-state in molecular frame of reference can be written as
\bea
 \mid \Psi_{gg}\rangle  = \frac{{ \mid( ^1S_0)\imath,\eta_{\imath}\rangle}^A{ \mid (^1S_0)\imath',\eta_{\imath}'\rangle}^B + { \mid (^1S_0)\imath',\eta_{\imath}'\rangle}^A { \mid (^1S_0)\imath,\eta_{\imath}\rangle}^B}{\sqrt{2}}  |I_g,\Phi_g\rangle_{\rm nucl}
\eea
where $A,B$ refer to the atoms and $ \mid I_g,\Phi_g\rangle_{\rm nucl}$ is the nuclear spin wave function with $\Phi_g$ the axial projection of total nuclear spin $I_g$, $\eta_{\imath}$ is the projection of the nuclear spin $\imath$ for the $^1S_0$ state. The electronic excited states can be written as
\bea
 \mid \Psi_{eg}\rangle &=& \frac{ \mid (^3P_1)f,\phi\rangle^A \mid (^1S_0)\imath,\eta_{\imath}\rangle^B + \mid (^1S_0)\imath,\eta_{\imath}\rangle^A \mid (^3P_1)f,\phi\rangle^B}{\surd 2}\\
 \mid \Psi_{ee}\rangle &=& \frac{ \mid (^3P_1)f,\phi\rangle^A \mid (^3P_1)f',\phi'\rangle^B + \mid (^3P_1)f',\phi'\rangle^A \mid (^3P_1)f,\phi\rangle^B}{\surd 2}
\eea
Here $\phi$ is the projection of the  total atomic angular momentum $f$ onto the molecular axis. $\vec F = \vec I + \vec J$ is the total molecular angular momentum, $\bf \Phi $ is the axial projection of $\vec F$. Here $\bf \Phi $ = $\phi^A + \phi'^B$ or $\bf \Phi $ = $\phi^A + \eta_{\imath}^B$, which serves as a good quantum number. $ ^{171}$Yb has nuclear spin $\imath = \frac{1}{2}$. At low energy, $s$-wave collision occurs for $I_g = 0$ and $p$-wave for $I_g = 1$ \cite{enomotoprl2008}. We consider low energy collision between a pair of spin-polarized $^{171}$Yb atoms inside cavity.

Here we extend the Movre-Pichlar model by including cavity photon states to form atomic and photonic product basis. The $\mid \uparrow \rangle$ refers to the state with $\phi$ = $\frac{1}{2}$, and $\mid \downarrow \rangle$ is used for $\phi$ = -$\frac{1}{2}$. For $\mid (^1S_0) \imath,\eta_{\imath}\rangle$ state, $\mid \uparrow \rangle$ is used for $\eta_{\imath}$ = $\frac{1}{2}$, and $\mid \downarrow \rangle$ is used for $\eta_{\imath}$ = -$\frac{1}{2}$. An atomic basis state is represented in the form $\mid \uparrow, \downarrow \rangle_{m n} = \mid \uparrow \rangle_{m} \otimes \mid \downarrow \rangle_{n} $ or $\mid \downarrow, \uparrow \rangle_{m n} = \mid \downarrow \rangle_{m} \otimes \mid \uparrow \rangle_{n} $ where the subscripts $m$ and $n$ stand for any two atomic level indexes $g, e_1, e_2$  or equivalently 0, 1, 2. The first arrow 
indicates the spin state of atom `$A$' while the second one denotes that of atom `$B$'. Now, at first we form seven symmetrised coupled atom-field states for $\bf \Phi$ = 0.
\begin{figure}
\includegraphics[width=\columnwidth]{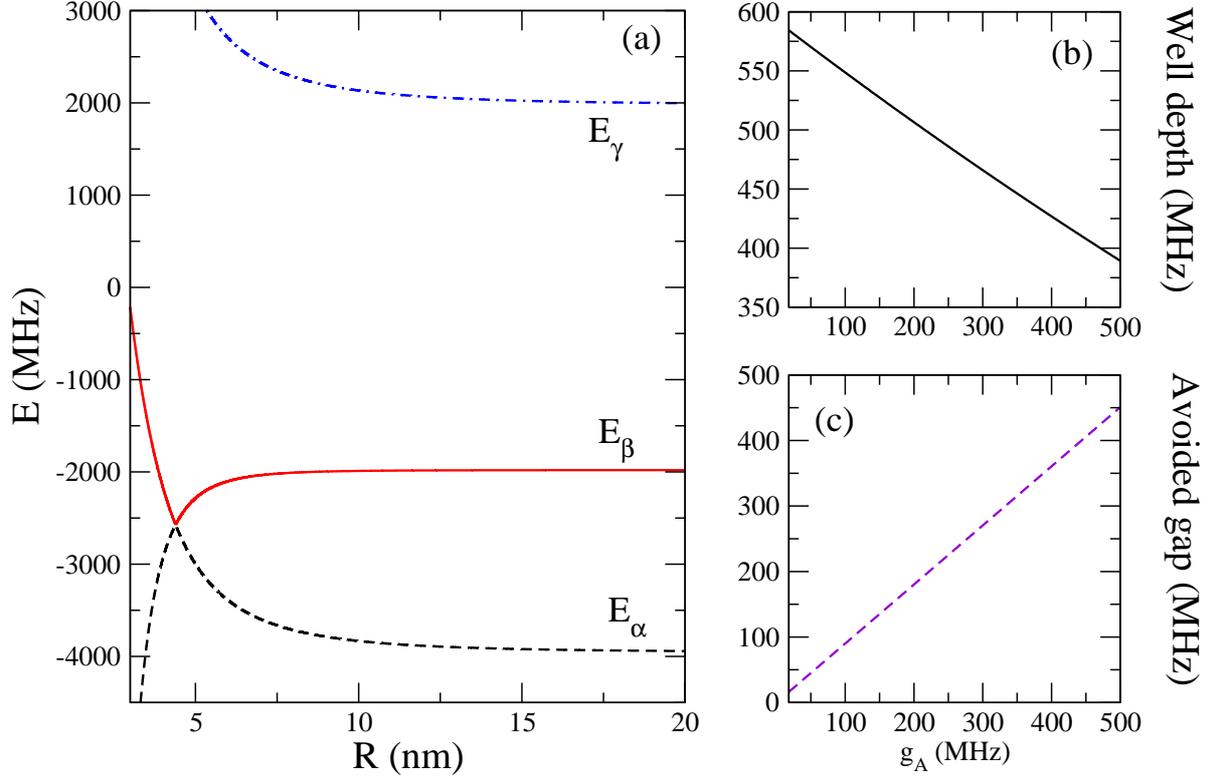}
\caption{(a) One-photon sector adiabatic potentials for $\rm {g}_A$ = 18 MHz, $\rm {g}_B $ = 0.8 $\rm {g}_A$ and $\delta_1 = -\delta_2$=-1.0 Mhz. Asymptotes correspond to states mentioned in Eq. (\ref{eq10}). The variation of well depth (b) and avoided gap (c) as a function of the coupling parameter. }
\label{fig3}
\end{figure}
\bea
\mid0; N_1,N_{2}+1 \rangle &=& \frac{1}{\sqrt{2}} \left ( \mid \downarrow, \uparrow \rangle_{e_1 g} + \mid \uparrow, \downarrow \rangle_{g e_1} 
\right ) \otimes|N_{{1}},N_{2}+1\rangle  \nonumber \\
\mid 1; N_1,N_{2}+1 \rangle &=& \frac{1}{\sqrt{2}}(\mid \uparrow, \downarrow \rangle_{e_1 g} + \mid \downarrow, \uparrow \rangle_{g e_1}) \otimes|N_{{1}},N_{2}+1\rangle  \nonumber\\
\mid 2; N_1,N_{2} \rangle &=& \frac{1}{\sqrt{2}} (\mid \downarrow, \uparrow \rangle_{e_2 e_1} + \mid \uparrow, \downarrow \rangle_{e_1 e_2} ) \otimes |N_{{1}},N_{{2}}\rangle \nonumber\\
\mid 3; N_1,N_{2} \rangle&=& \frac{1}{\sqrt{2}} (\mid \uparrow, \downarrow \rangle_{e_2 e_1} + \mid \downarrow, \uparrow \rangle_{e_1 e_2} ) \otimes|N_{{1}},N_{{2}}\rangle \nonumber\\
\mid 4;N_1+1,N_{2}+1 \rangle &=& \frac{1}{\sqrt{2}}(\mid \downarrow, \uparrow \rangle_{g g} + \mid \uparrow, \downarrow \rangle_{g g} ) \otimes|N_{1}+1,N_{2}+1\rangle\nonumber\\
\mid 5; N_1+1,N_{2} \rangle &=& \frac{1}{\sqrt{2}}(\mid \downarrow, \uparrow \rangle_{e_2 g}+ \mid \uparrow, \downarrow \rangle_{g e_2} ) \otimes|N_{1}+1,N_{{2}}\rangle \nonumber\\
\mid 6; N_1+1,N_{2} \rangle &=& \frac{1}{\sqrt{2}}\mid \uparrow, \downarrow \rangle_{e_2 g} + \mid \downarrow, \uparrow \rangle_{g e_2} ) \otimes|N_{1}+1,N_{{2}}\rangle
\label{eq7}
\eea
where $N_{1(2)}$ is the number of photons in field-mode 1(2). Now, the states written in Eq.(\ref{eq7}) are Bell basis of atomic states. PLR or quasi-molecular interactions allow these maximally entangled states to arise quite naturally in the dynamics.

\subsection{Symmetric and antisymmetric dressed basis}
The atom-field basis states in Eq.(\ref{eq7}) appear to be not very convenient to reveal a physically intuitive picture. In Eq.(\ref{eq7}) the states $\mid 0\rangle$ and $\mid 1 \rangle$ have same photonic state $\mid N_1, N_2+1 \rangle$, but the atomic states are different, they are actually spin flip states, which are  degenerate in energy. Similarly, the pair of states $\mid 2\rangle$, $\mid 3 \rangle$ have same photonic state $\mid N_1, N_2 \rangle$, but the atomic states are different but degenerate. Likewise, $\mid 5\rangle$, $\mid 6 \rangle$ have same photonic state $\mid N_1+1, N_2 \rangle$, but the atomic states are spin flipped and degenerate. Hence, energetically degenerate dressed states in Eq.(\ref{eq7}) need to be expressed in symmetric and antisymmetric combination. We denote the new basis in the form
\bea
\mid a \rangle &\equiv& \mid a; N_1,N_{2}+1 \rangle = \frac{1}{\sqrt{2}} (\mid 0 \rangle - \mid 1\rangle)\nonumber\\
\mid b \rangle &\equiv& \mid b; N_1,N_{2}+1 \rangle = \frac{1}{\sqrt{2}} (\mid 0 \rangle + \mid 1\rangle)\nonumber\\
\mid c \rangle &\equiv& \mid c; N_1,N_{2}\rangle = \frac{1}{\sqrt{2}} (\mid 2\rangle + \mid 3\rangle)\nonumber\\
\mid d \rangle &\equiv&  \mid d; N_1,N_{2} \rangle = \frac{1}{\sqrt{2}} (-\mid 2 \rangle + \mid 3\rangle) \nonumber\\
\mid f \rangle &\equiv& \mid f;N_1+1,N_{2} \rangle = \frac{1}{\sqrt{2}} (\mid 5 \rangle + \mid 6\rangle)\nonumber\\
\mid g \rangle &\equiv& \mid g; N_1+1,N_{2}\rangle = \frac{1}{\sqrt{2}} (-\mid 5 \rangle + \mid 6\rangle)\nonumber\\
\mid e \rangle &\equiv& \mid e; N_1+1,N_{2}+1\rangle = \mid 4 \rangle 
\label{eq9}
\eea
\begin{figure}
\includegraphics[width=\columnwidth]{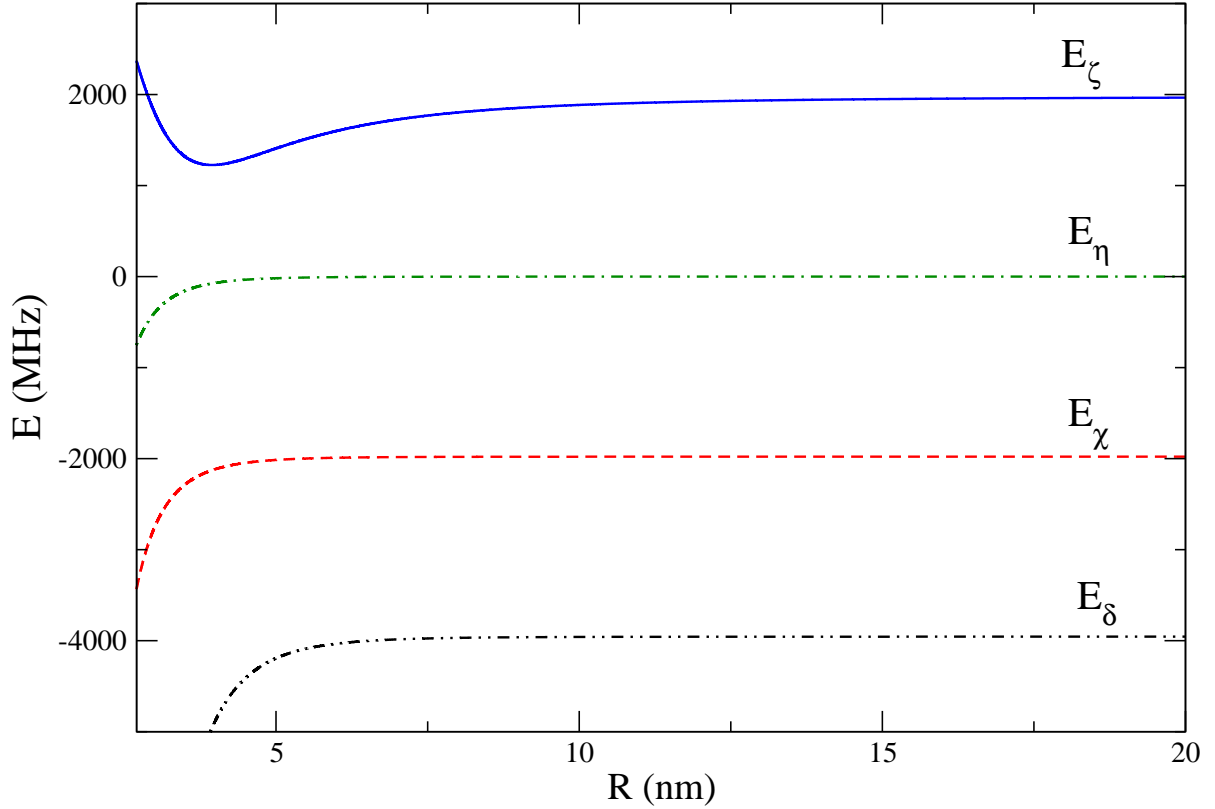}
\caption{Two-photon sector adiabatic potentials for $\rm {g}_A$ = 18 MHz, $\rm {g}_B $ = 0.8 $\rm {g}_A$ and $\delta_1 = -\delta_2$=-1.0 Mhz. Asymptotes correspond to states mentioned in Eq. (\ref{eq10}).}
\label{fig2}
\end{figure}
These states are superposition of Bell states of Eq.(\ref{eq7}). We can discuss two types of Bell states, one results from mutual spin-flip in two different electronic states and the other from spin flip in the same electronic state. These basis states may be viewed as superposition of either type. For simplicity of our analysis we consider that each cavity mode has only a single photon. In these basis states the hamiltonian becomes block-diagonalised where $\mid b\rangle, \mid c\rangle, \mid g\rangle$ forming block-A, we name it one-photon sector and  $\mid e\rangle, \mid a\rangle, \mid d\rangle, \mid f\rangle$ another block-B or two-photon sector.

\subsection{Fano effect in CQED}
We consider nonadiabatic interaction due to pseudo-crossing as described in Appendix-B. Nonadiabatic effects give rise to the coupling between a bare continuum and a bound state, leading to the formation of a new dressed continuum that can be treated by Fano's theory \cite{fanopr1961}. Under semi-classical approximation, the nonadiabatic effects due to pseudo-crossing is normally described by Landau-Zenner-Stueckelberg theory \cite{LZS}.  We here show that one can develop an alternative approach using Fano's method when the upper potential is a binding potential that can support at least one bound state.  The essence of this method is to diagonalize the system including nonadiabatic coupling, and then use the eigenvalues and eigenstates to calculate the dynamics. The connection  with the traditional Landau-Zenner-Stueckelberg approach can be established by calculating time-dependent transition probabilities.  
  
Let us consider two dressed-state adiabatic potentials near the pseudo-crossing either in one- or two- photon sector as shown in Fig.\ref{fig3} and Fig.\ref{fig2}. In both sectors we consider one continuum of states of relative motion between one ground-state ($^1S_0$) and the other excited-state ($^3P_1$) atoms. This continuum interacts with bound state of the upper binding potential.  In Fig.\ref{fig3}, the upper potential ($E_{\beta}$) has a well allowing it to support two-atom bound states. The lower one ($E_{\alpha}$) has a barrier near the avoided crossing and so it is anti-binding, leading to the free states (scattering) of the two atoms. In Fig.\ref{fig2}, the upper potential $E_{\zeta}$ can support bound states. 

We here restrict our discussion to two-channel model problem, where the two channels concerned are the asymptotic cavity-dressed states that correspond to the two dressed potentials showing the pseudo-crossing. For simplicity, we consider that a molecular bound state $\mid b \rangle$ supported by the upper binding potential in one- or two-photon sector is coupled to the continuum of scattering states $\mid  a_{E'}\rangle$ with collision energy $E'$ in the lower potential via nonadiabatic interaction. The hamiltonian ($\hat{W}$) can be written as
\bea
\hat{W} &=& \hat{W}_0 + \hat{W}'\nonumber\\
\hat{W}_0 &=& W_{b} \mid b \rangle \langle b \mid + \int E' dE' \mid a_{E'} \rangle \langle a_{E'} \mid\\
\hat{W}' &=& \int dE' \Lambda_{a}^{b} (E') \mid b \rangle \langle a_{E'} \mid + c.c.
\eea
where $W_b$ is the binding energy of the bound state $\mid b \rangle$. This interaction leads to the formation of energy normalized dressed state $\mid E \rangle$
\begin{equation}
  \mid E\rangle =  A_{E} \mid b \rangle + \int dE' C_{E'}(E) \mid  a_{E'}\rangle
 \end{equation}
Where $A_{E}$ and $C_{E'}(E)$ are the dressed amplitudes. From the time-independent Schr$\textrm{\"{o}}$dinger equation $\hat{W} \mid E\rangle = E \mid E\rangle $ we obtain following set of coupled differential equations:
\bea
(W_b -E) A_E &=& -\int dE' \Lambda_a^b (E') C_{E'}\nonumber\\
(E'-E) C_{E'} (E)&=& - \Lambda_b^a (E')A_E 
\label{eq20}
\eea
where $\Lambda_a^b = \langle b \mid W' \mid a \rangle$. Solving Eq.(\ref{eq20}) we obtain after some algebra
\bea
A_E &=& \frac{\Lambda _a^b (E)}{E-W_b + i \frac{\Gamma}{2}}\nonumber\\
C_{E'} (E) &=& \frac{\Lambda_b^a(E')}{E-E'} A_E + \delta (E-E')
\label{eq21}
\eea
where $\Gamma = 2\pi \mid \Lambda_a^b (E) \mid^2$ is the  width of the bound state due to nonadiabatic coupling. Following Ref.\cite{debpra2012}, we obtain the $T$ matrix element 
\be
  T = -e^{i\eta_{bg}} {\rm sin}\hskip.05cm \eta_{bg} \hskip.1cm + \pi  A_E \Lambda^{a}_{b} (E) e^{2i\eta_{bg}}
  \label{eq22}
\ee
Where $\eta_{bg}$ is the background (without the nonadiabatic coupling) phase shift. The elastic scattering cross section is
 \be
 \sigma_{el} = \frac{4\pi }{k^2} \mid T \mid^2
  \label{eq26}
 \ee 
 
 It is to be noted here that the time-dependent wave function can be formed by 
 \bea 
\mid \Psi(t)\rangle  = \int dE' \exp[-i E' t/\hbar] \mid E' \rangle \langle  E' \mid \Psi(0)  \rangle 
 \eea 
The evaluation of energy integration in Eq.(20) is not possible analytically. So, one has to resort to numerical evaluation of this integral. Note that the states $\mid b \rangle$ and $\mid a_E \rangle$ in Eq.(15) belong to two different channels or internal states of diatom-photon hybrid system. So, the time-dependent wave function will be a superposition of different internal states which will evolve in time. The initial state $\Psi (0)$ is a product of internal and external motional state of the two atoms. Equation (20) describe the coherent evolution of the system.

\section{Results and Discussion}
In our numerical calculations we consider $p$-wave collision between a pair of spin polarized fermionic $^{171}$Yb atoms in strong-coupling CQED regime. We are here considering $^1S_0$ - $^3P_1$ intercombination transition, which has a narrow line width 182 KHz \cite{ueda:pra:2010}, hence the atomic decay $\gamma$ is too small and can be safely neglected in this coupling regime. In a recent experiment using $^{171}$Yb in two-mode cavity QED setup, the cavity decay $\kappa$ is found to be 4.8 MHz \cite{ueda:prl:2011}, though the strong coupling regime is not not achieved there. We take coupling constant $\rm{g_{A}}, \rm{g_{B}} \backsimeq$  18 MHz (Kimble's group has achieved strong coupling regime of CQED with Cs atoms. We take our coupling constants comparable to experimental ones\cite{kimble:jpb:2005}). The relevant parameters for $^{171}\rm Yb$ are $a_{\rm hf}$ = 3957 MHz \cite{clarkpra1979}, long-range dispersion coefficients $C_6 = 1932 $ a.u. \cite{porsevpra2014} and $C_6=2810$ a.u. \cite{enomotoprl2008}
 for molecular potentials asymptotically corresponding to $^1S_0$ + $^1S_0 $ and $^1S_0$ + $^3P_1$ separated atoms, respectively. $V_{QQ}(R)$ is the quadrupole interaction when both atoms are in excited state $^3P_1$ + $^3P_1$. More technical details about this quadrupole interaction are given in Appendix-A.

\begin{figure}
\includegraphics[width=\columnwidth]{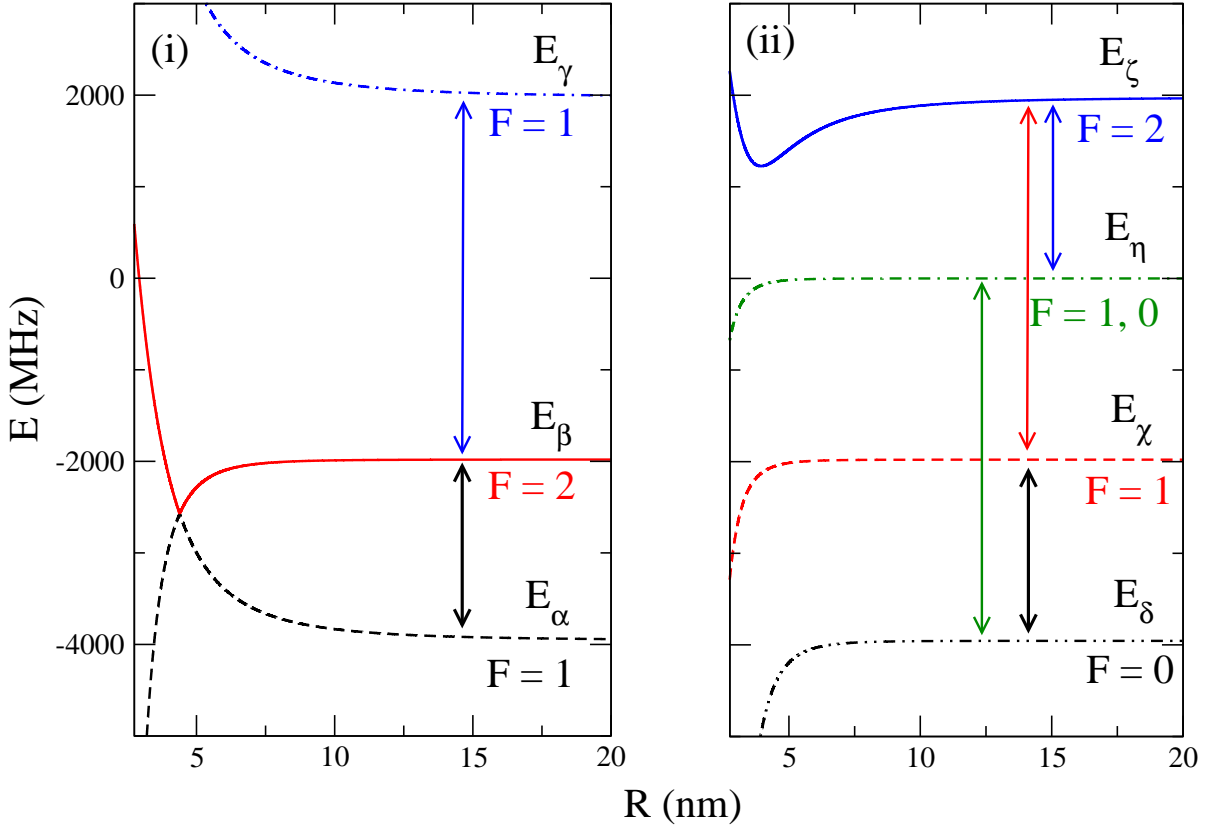}
\caption{Asymptote potentials of one-photon (i) and two-photon (ii) sector. Selection rules on total angular momentum F make these two sectors to be independent of one another. Arrows show the dipole allowed transition pathways.}
\label{fig7}
\end{figure}
We have diagonalised the hamiltonian of Eq.(\ref{eq2}) by using $\mid a \rangle, \mid b \rangle,\mid c \rangle ...\mid g \rangle$ of Eq.(\ref{eq9}) and we find block diagonalised eigen potentials. Three of them $E_\alpha, E_\beta, E_\gamma$ (named according to increasing energy) belong to one-photon sector as displayed in Fig.\ref{fig3} and rest four of them $E_\delta, E_\chi, E_\eta, E_\zeta$ belong to two-photon sector as shown in Fig.\ref{fig2}. The corresponding eigenstates can be named as $\mid \alpha \rangle, \mid \beta \rangle, \mid \gamma \rangle,\mid \delta \rangle,  \mid \chi \rangle, \mid \eta \rangle, \mid \zeta \rangle$. These states asymptotically correspond to bare dressed states of Eq.(\ref{eq9}), that is 
\bea
 \mid \alpha \rangle &(R \rightarrow \infty)\longrightarrow& \mid b\rangle,\hskip .5cm \mid \beta \rangle (R \rightarrow \infty)\longrightarrow \mid c\rangle,\nonumber\\
 \mid \gamma \rangle &(R \rightarrow \infty)\longrightarrow& \mid g\rangle,\hskip .5cm \mid \delta \rangle (R \rightarrow \infty)\longrightarrow \mid a\rangle,\nonumber\\
 \mid \chi \rangle &(R \rightarrow \infty)\longrightarrow& \mid d\rangle,\hskip .5cm \mid \eta \rangle (R \rightarrow \infty)\longrightarrow \mid e\rangle,\nonumber\\
\mid \zeta \rangle &(R \rightarrow \infty)\longrightarrow& \mid f\rangle
\label{eq10}
\eea
\begin{figure}
\includegraphics[width=\columnwidth]{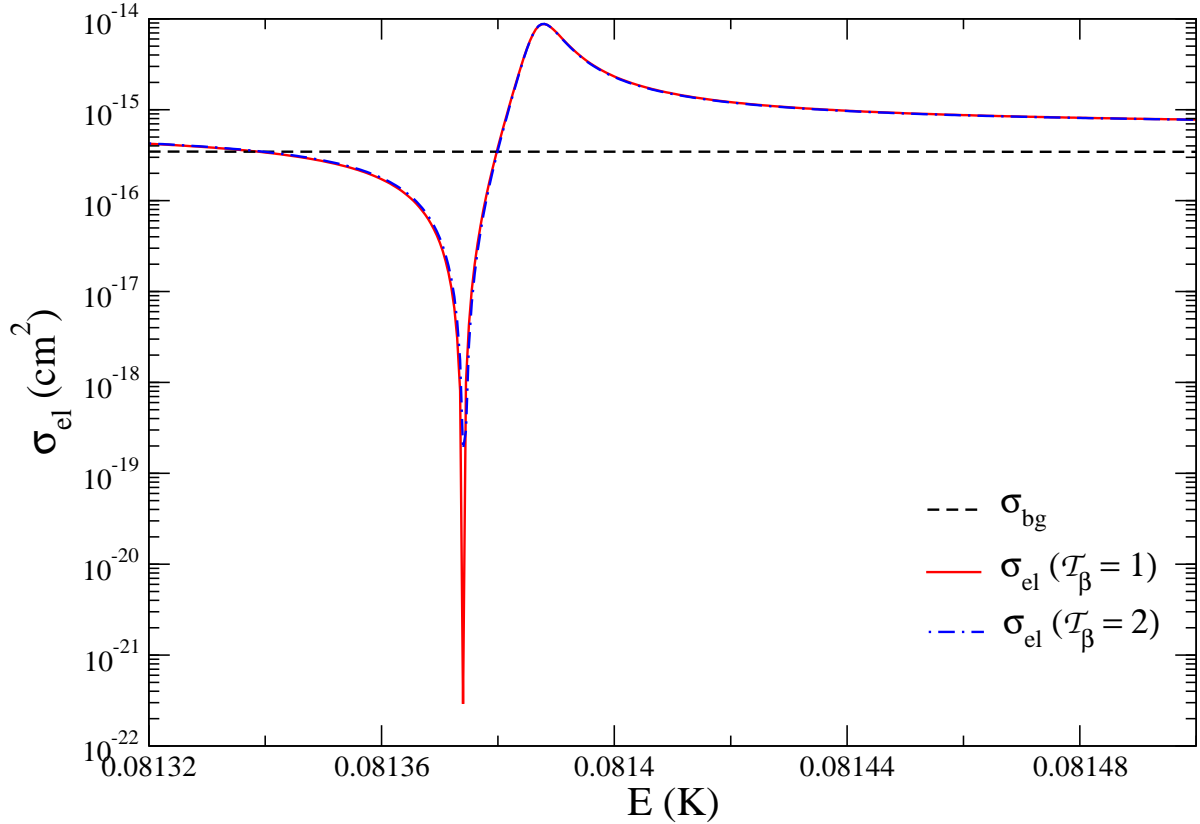}
\caption{The scattering cross section $\sigma_{el}$ for $\mathscr{T}_{\beta}$ = 1 (solid) and $\mathscr{T}_{\beta}$ = 2 (dash-dotted) are plotted as a function of collision energy in Kelvin. The background scattering cross section $\sigma_{bg}$ (dashed) is also shown.}
\label{fig4}
\end{figure}

 Asymptote analysis shows that according to the $\Delta$F selection rule, $\Delta F = 0, \pm 1$, the eigen potentials become block diagonalized. In Fig.\ref{fig7} one can notice that each block has only the selection rule allowed dipole transition pathways. We name the one-photon sector as in that sector at most one photon is present in the bare  states ($\mid b \rangle$, $\mid g \rangle$), and in two-photon sector at most two photons are present in the bare  state ($\mid e \rangle$).  

\Table{\label{abrefs}Bound state energies (in MHz) of potential $E_{\beta}$ (in one-photon sector) for different coupling parameter $\rm {g}_A$ (in MHz). Vibrational number ($v$) represents number of nodes present in the bound-state wave function.} 
\br
$\mathscr{T}_{\beta}$ & v & $E_v$ & $E_v$ & $E_v$\\
 &  & ($\rm {g}_A$ = 18 MHz) & ($\rm {g}_A$ = 100 MHz) & ($\rm {g}_A$ = 500 MHz)\\
\mr
1 & 0 & -281.5 & -274.9 & -201.6\\
1 & 1 & -53.8 & -52.2 & -31.2\\
1 & 2 & -1.9 & -1.7 & \\
\br
\label{tb1}
\endTable
The potential ($E_{\zeta}$) having depth about 750 MHz appears in two-photon sector, it remains approximately unchanged with moderate changes in cavity parameters. Another binding potential ($E_{\beta}$) appears in one-photon sector which asymptotically corresponds to $^3P_1$ + $^3P_1$ states, i.e. $\mid \beta \rangle$ state of Eq.(\ref{eq10}). Unlike those in two-photon sector, the dressed potentials in one-photon sector are highly sensitive to the tuning of cavity parameters. Quadrupole interaction plays an important role in the appearance of this binding potential in one-photon sector. In free space the avoided crossing will transform into a crossing and thus this new binding potential does not appear in the absence of field-dressing. We notice that this binding potential which can support a few bound states can be modified by changing the cavity coupling parameter as shown in Table \ref{tb1}. We calculate the bound-state wave function and energies using the standard Numerov algorithm.

Next, we perform scattering calculations again using Numerov algorithm. We consider two-channel model in one-photon sector with the channel being $\alpha$ and $\beta$. The channel $\gamma$ is far away from the two channels and so it has practically no influence on low energy dynamics near pseudo-crossing. We consider $\alpha$ as incident channel which at short range corresponds to 0$^+_u$ potential. At short range $\textrm{Yb}_2$ has simple electronic state having $^1S_0$ symmetry and therefore the molecular ground state potential is of $^1\Sigma_g$ molecular symmetry with no electronic orbital and spin quantum number. 
\begin{figure}
\includegraphics[width=\columnwidth]{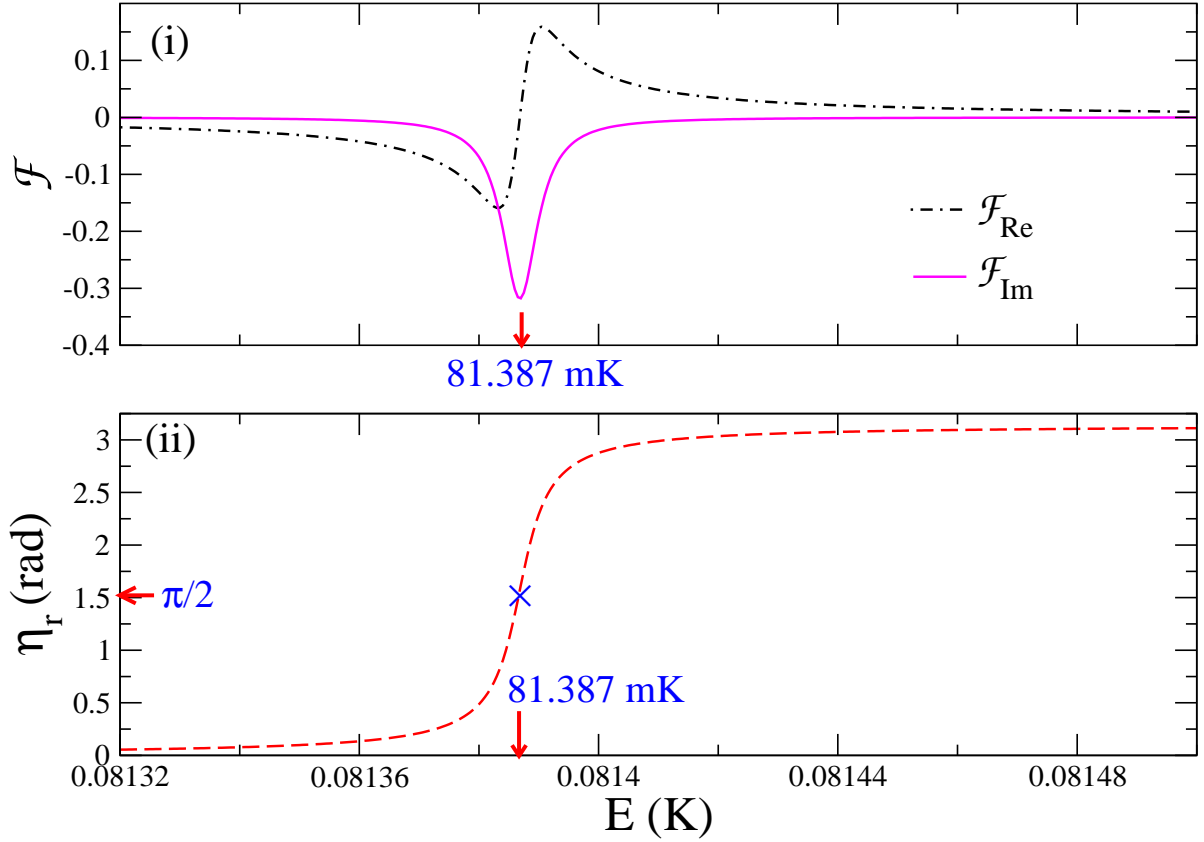}
\caption{The top panel (i) shows the variation of ${\cal F}_{Re}$ and ${\cal F}_{Im}$ with Energy and the bottom panel (ii) shows the variation of $\eta_r$ with energy.The crossmark and arrowheads denote the position of resonance where $\eta_r = \frac{\pi}{2}$ at collision energy 81.387 mK.}
\label{fig6}
\end{figure}
As we are using intercombination transitions which are dipole allowed but spin forbidden, the short range of excited state of $^1S_0$ + $^3P_1$ should correspond to 0$^+_u$ molecular state. We consider the radial coupling between the individual channel solutions: for $\alpha$ channel we consider scattering eigenfunction, bound state for $\beta$ channel. Now, we follow Fano theory to form the energy-normalized dressed eigenfunction $\mid E \rangle$ of the system. The T matrix element of Eq. (\ref{eq22}) can be rewritten as
\bea
T &=& -e^{i\eta_{bg}} {\rm sin}\hskip.05cm \eta_{bg} \hskip.1cm + \pi {\cal F} e^{2i\eta_{bg}}\nonumber\\
&=& T_{bg} + e^{2i\eta_{bg}} T_r
\eea
where $T_{bg} = -e^{i\eta_{bg}} {\rm sin}\hskip.05cm \eta_{bg} \hskip.1cm $ is the $T$-matrix element in the absence of nonadiabatic coupling. $T_r$ is the effect of  nonadiabatic coupling which essentially couples the scattering and the bound state. $T_r = -e^{i\eta_{r}} {\rm sin}\hskip.05cm \eta_r$ ,$\hskip.1cm$ where $\eta_r = \textrm{tan}^{-1}\left(\frac{{\cal F}_{Im}}{{\cal F}_{Re}}\right)$. Here ${\cal F}_{Re}, {\cal F}_{Im}$ are the real and imaginary part of ${\cal F} =  A_E \Lambda^{a}_{b} (E)$.  

In Fig.\ref{fig4} we have plotted $\sigma_{el}$ of Eq.(\ref{eq4}) of  $\mathscr{T}_{\beta} = $ 1, 2 for different collision energies. Here $\mathscr{T}_{\alpha(\beta)} = \textbf{F}_{\alpha(\beta)} + \vec{\ell} $ is the total angular momentum of the corresponding channel $\alpha (\beta)$, where $\vec{\ell}$ denotes the angular momentum of the relative motion between the two atoms. For $^3P_1$ + $^3P_1$ asymptote i.e.  ${\beta}$ channel, $\textbf{F} = 2$ and $\ell = 1$, so $\mathscr{T}_{\beta}$ = 1, 2, 3. On the other hand for $^1S_0$ + $^3P_1$ asymptote  i.e. ${\alpha}$ channel, $\textbf{F} = 1$ and $\ell = 1$, so $\mathscr{T}_{\alpha}$ = 0, 1, 2. We take $\mathscr{T}_{\alpha} = 1, 2$ of $ {\alpha}$ channel being coupled to bound vibrational states of potential $ {\beta}$ with $\mathscr{T}_{\beta} =$ 1,2 respectively with $M_{\mathscr{T}_{\beta}} = M_{\mathscr{T}_{\alpha}}$ via non adiabatic coupling. We find that there is a Fano resonance at about 81.38 mK (approximately) in scattering cross section for both 
$\mathscr{T}_{\beta}$ = 1, 2. The resonance is more prominent for$\mathscr{T}_{\beta}$ = 1 than that in case of $\mathscr{T}_{\beta}$ = 2. The bound state energy $E_b$ ($\mathscr{T}_{\beta}$ = 1) is about 1695.83 MHz or 81.387 mK above the scattering threshold. Hence this resonance is clearly associated with the presence of bound state $v = 0$, $\mathscr{T}_{\beta} = 1$ there. The background phase shift $\eta_{bg} \approx$ 6.008$^c$, remains approximately constant in this energy regime. To have a close look at the resonance, we have plotted ${\cal F}_{re}$, ${\cal F}_{Im}$ and $\eta_r$ for $\mathscr{T}_{\alpha} = \mathscr{T}_{\beta} = \mathscr{T} = 1$ in Fig.{\ref{fig6}}. We can detect a clear resonance at about 81.387 mk as $\eta_r$ changes sharply through $\frac{\pi}{2}$ near this energy. Analytically,  we find that at $\eta_{bg} = 2\pi - \eta_r$, $T$-matrix element vanishes, i.e the elastic scattering cross-section attains a minimum. We find that at $E$ = 81.374 mK, $\eta_r = 0.2747^c$ and $\eta_{bg} = 
6.0084^c$, it is equal to the value of background phase shift at that collision energy and hence the elastic scattering cross section diminishes. The background and resonance scattering amplitudes can interfere destructively or constructively, leading to the minimum or maximum in $\sigma_{el}$ vs. energy plot, respectively at $E$ = 81.374 mK and $E$ = 81.387 mK . The peak position appears near the energy of the bound state. This can be regarded as nonadiabatic coupling-induced Fano-Feshbach resonance.

We have also performed the similar calculation for two-photon sector (Fig.{\ref{fig2}) taking $\zeta$ as upper channel having binding PLR potential and $\eta$ as lower scattering channel for $\mathscr{T} = 1$. The upper binding potential supports as many as 8 bound states. Without cavity coupling the binding energies match well with the reported free space values \cite{enomotoprl2008}. We have found that with strong coupling they do not change appreciably. We further find that the nonadiabatic coupling is smaller here, because the derivative of the potential surface varies smoothly with $R$. Hence, although  we find a  resonance at the corresponding bound-state energy, but it is not prominent and elastic scattering cross section changes by one order of magnitude only, whereas in one-photon sector it changes by 7 orders of magnitude causing a sharp resonance there. 

We have done all the above calculations for weak coupling regime, where we have set $\rm g_A$ = 2.8 MHz which is the same as in the experiment by Takeuchi {\it et al.} \cite{ueda:pra:2010}. This is lesser than cavity decay constant $\kappa$. The results in the weak-coupling regime are qualitatively similar to those found for strong coupling case. Only due to the change of coupling parameter the aforesaid bound eigen potential and the scattering potential in one-photon sector are qualitatively different. Hence the bound state energy changes to -281.8 MHz for $v$= 0 and $\mathscr{T} = 1$. We have found the resonance in this weak coupling case at 1695.43 MHz. We have checked the case with much stronger coupling $\rm g_A$ =  50 MHz, we found the resonance there at 1697.96 MHz.

\section{Conclusions}

In conclusion we have presented a dressed state formalism for the treatment of two slowly colliding V-type three-level atoms interacting  with a two-mode quantized cavity field;  and shown that it is possible to generate Fano-Feshbach resonances in cavity QED regime. For the sake of simplicity, in our formalism, we have used coupled diatomic and two-mode photonic states with each mode containing either zero (vacuum) or one photon. We have illustrated our numerical results for a pair of fermionic $^{171}$Yb atoms in the cavity. The atom-pair collectively interacts with the cavity modes predominantly at nanoscale separations which are quite long-ranged compared to that of typical molecular interactions. The adiabatic dressed states are shown to belong to two separate sectors of interactions - one involving only a single photon in either mode and the other one involving two photons. We have identified one prominent pseudo-crossing point between two adiabatic potentials in the one-photon 
sector. The nature of this  crossing point is shown to depend strongly on the atom-field coupling strengths.

We have shown how the non-adiabatic coupling near this pseudo-crossing leads to Fano effect in the intra-cavity scattering between the ground and excited atoms. In general,there is a possibility of cavity-photon mediated long-range interaction \cite{ddi,Zheng:2000,
Petrosyan:2003, Esslinger:Science:2012, Ritsch:rmp:2013, Blatt:prl:2013, reimannprl2015} between atoms. This occurs due to transitions at a single-atom level. But this kind of interaction is of much longer range than the nanoscale molecular separation which we studied in this paper. There is a way out to avoid the photon-mediated interaction inside the cavity such that one can explore only molecular interaction.This is possible if the free bound transition frequency is far below the threshold of the excited potential. Then, transitions at a single atomic level can be neglected when the cavity field is tuned near the free-bound transition.Since in our model calculations we have considered only two-atom case, and not considered the many-particle case, mean-field effect does not arise in our case.

The resonances we studied here are of Fano-type showing asymmetric profile. The minimum of the scattering cross-section represents the effect of quantum interference. At this Fano minimum, the decay of the bound state to the continuum is highly suppressed implying that the bound state is long-lived when the energy of the system is tuned near the Fano minimum. In this work we have provided a proof-of-principle of Fano effects in CQED and the effect of cavity fields on the possibility of manipulating molecular interactions between ground- and long-lived excited-state atoms. To the best of our knowledge, experimentally no such study has so far been attempted. In fact, coherent collisions involving excited-state atoms have not been studied much. Thanks to the progress in high precision photoassociation (PA) spectroscopy, exploration of such collisions have now become possible due to the accessibility of meta-stable excited states by PA spectroscopy \cite{takahashi2016}. We hope that such coherent spectroscopic tools will be extended to CQED in near future.

Finally, this study is important for an effective optical Feshbach resonance between ground- and excited-state atoms. The interactions between cold atoms is tunable by a magnetic  \cite{chinrmp2010} or optical \cite{fedichev1996} or magneto-optical Feshbach resonance \cite{deb:jpb:2010}. The major hindrance to efficient manipulation of atom-atom interactions by an optical method results from the spontaneous emission from the molecular bound states. In the strong-coupling CQED regime, one can ignore the atomic or molecular relaxation processes. Therefore cavity-coupling to molecular states can provide an alternative route for an efficient optical Feshbach resonance. Our results suggest that it is possible to devise a cavity-based method for manipulation of  interatomic interactions.

\section*{Acknowledgment}
One of us (BD) gratefully acknowledges many helpful discussions with Professor Gershon Kurizki, Weizmann Institute of Science.

\appendix
\section{Quadrupole Interaction}
The quadrupole interaction $V_{QQ}(R)$ is the first order correction to molecular term.
\bea
V_{QQ}(R) &=& \frac{Q^{11}|e^A_1\rangle\langle e^B_1|+Q^{12}|e^A_1\rangle\langle e^B_2|+Q^{21}|e^A_2\rangle\langle e^B_1|+Q^{22}|e^A_2\rangle\langle e^B_2| + h.c.}{R^5}
\eea
with $Q_{ij}$ being quadrupole-quadrupole interaction coefficients given by
\bea
Q^{ij} &=& \sum^{2}_{q=-2} (\frac{4!}{(2-q)!(2+q)!}) (F_q)_A (F_{-q})_B
\eea
The first order correction to molecular energies correlating to two $^3P_1$ states comes from quadrupole-quadrupole contribution. $F_q$ is the quadrupole spherical tensor \cite{dereviankoprl2001}.
\bea
F_q = -|e|\sum_i r^2_iC^{(2)}_{q}(\hat r_i)
\eea
where summation is over atomic electrons , $\hat r_i$ is the position vector of electron i and $C^{(2)}_{q}(\hat r_i)$ are the normalized spherical harmonics. The quadrupole moment of $^3P_1$ atomic state, defined conventionally as
\bea
\wp = 2\langle^3P_1,M_J=1|F^2_0|^3P_1,M_J=1\rangle = \sqrt{\frac{2}{15}} \langle^3P_1||F||^3P_1\rangle
\eea
where $\langle^3P_1||F||^3P_1\rangle$ is the reduced matrix element of the tensor which is -19.7 a.u.\cite{Buchachenkoepjd2011}. Actually $^3P_1$ is not pure spin-orbit LS coupling, but a mixture of higher lying $^1P_1$ state such that\cite{junyepra2007},
\bea
\mid^3P_1\rangle = \alpha \mid^3P^0_1 \rangle + \beta \mid^1P^0_1 \rangle
\eea
For $^{171}\rm Yb, {\alpha}^2$ = $(\pm0.99159)^2 >\,> \beta^2 = (\pm 0.12939)^2$. Hence we are taking only contribution of $ \mid ^3P^0_1\rangle$ for calculating quadrupole interaction. In general, the quadrupole moment matrix element can be written as \cite{dereviankoprl2001}
\bea
\langle^3P_1,M_J|F^2_0|^3P_1,M'_J\rangle &=& \langle J,M_J|F^2_0|J,M'_J\rangle\\\nonumber 
&=& (-1)^{(J-M_J)} \langle J||F||J\rangle \threej{J}{2}{J}{-M_J}{0}{M'_J}
\eea
Quadrupole interaction makes the binding potential in one-photon sector deeper. For $^3P_0$ + $^3P_0$ the second order energy correction is  $C_6 = 3886$ a.u.\cite{porsevpra2014}, we have taken same $C_6$ for molecular potential that asymptotically correspond to $^3P_1$ + $^3P_1$ separated atoms.

\section{Non-adiabatic Interaction}
\begin{figure}
\includegraphics[width=\columnwidth]{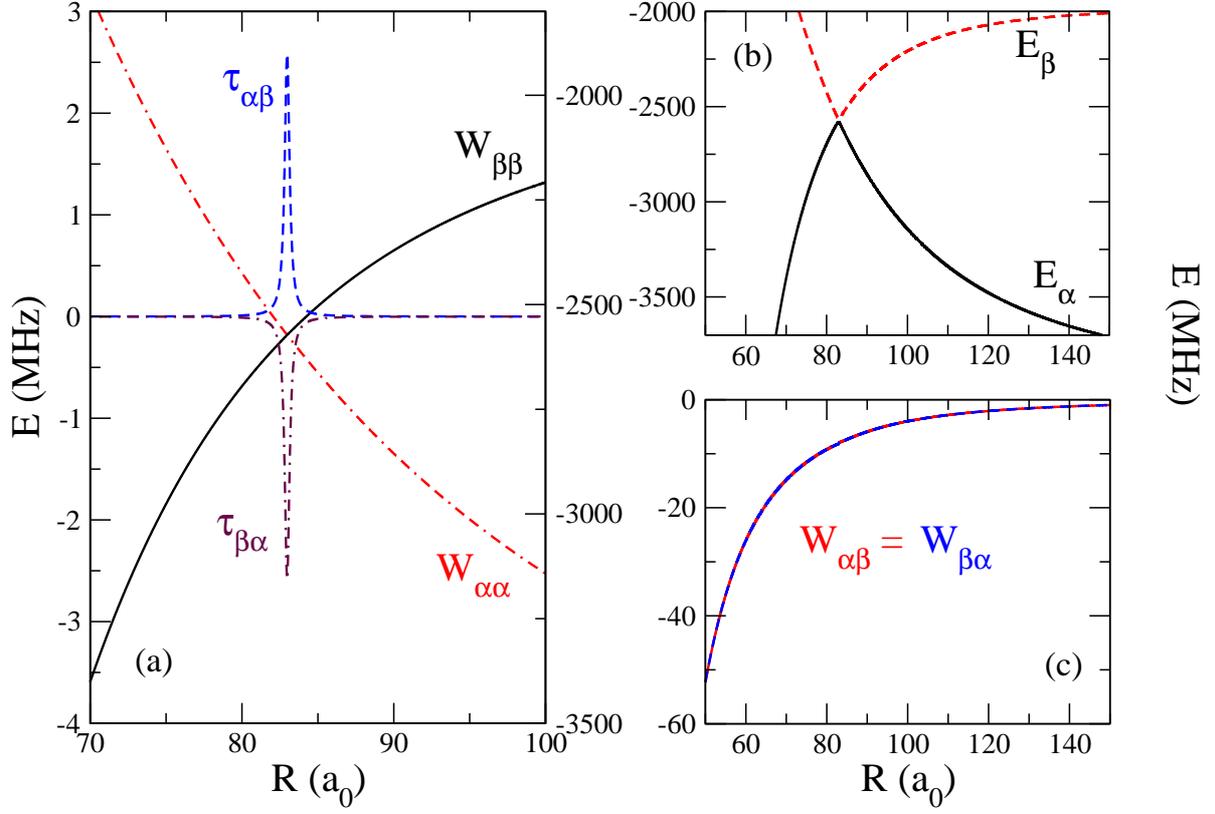}
\caption{(a) Diabatic potential picture for one-photon sector with potential $E_{\alpha}$ and $E_{\beta}$. Diagonal W-matrix elements show an avoided crossing. $\tau_{ij}$'s are the nonadiabatic coupling matrix elements, (b) the avoided crossing of $E_{\alpha}$ and $E_{\beta}$ in adiabatic case, (c) Off-diagonal W-matrix elements.}
\label{fig5}
\end{figure}
 In the potential coupling picture the total wave function can be expressed in terms of adiabatic normalized eigen function $\phi$ of $H_{ad}$. We thus have
\bea
H_{\rm ad} \phi(R) &=& \hbar \widetilde{\omega} \phi(R)\nonumber\\
\Psi(R) &=& \sum_{i=1,2}( \chi_i\phi_i)
\label{eq12}
\eea
where $\widetilde{\omega}$ is the adiabatic eigen energy matrix. Inserting (\ref{eq12}) into Schr$\rm{\ddot{o}}$dinger equation
\bea
\left[-\frac{\hbar^2}{2\mu} \left(\frac{d^2}{dR^2} - \frac{l(l+1)}{\hbar^2 R^2}\right) + H_{ad}\right]\Psi &=& E\Psi
\eea
we get
\bea
\left(\frac{d^2}{dR^2} + 2 \tau^{(1)} \frac{d}{dR} + \tau^{(2)}\right)\chi - \left[\frac{l(l+1)}{\hbar^2 R^2} + \frac{2\mu}{\hbar^2}(\hbar\widetilde{\omega} -E )\right ]\chi = 0
\label{eq14}
\eea
Where $\tau^{(1)}$ is an antisymmetric and $\tau^{(2)}$ is a symmetric matrix with the elements
\bea
\tau^{(1)}_{\rm \mu \nu} (R) = \phi^{*}_{\rm \mu}(R) \frac{d}{dR} \Phi_{\rm \nu} (R) = -\tau^{(1)}_{\rm \nu \mu} (R)\nonumber\\
\tau^{(2)}_{\rm \mu \nu} (R) = \phi^*_{\rm \mu}(R)\frac{d^2}{dR^2} \Phi_{\rm \nu} (R)
\eea
$\tau^{(1)}_{\rm \mu \nu}$ is usually called as nonadiabatic coupling matrix element. The higher order terms $\tau^{(2)}_{\rm \mu \nu} (R)$ are dropped. For sake of convenience we have transformed Eq.(\ref{eq14}) into the following form \cite{child,baerjcp1980}
\be
\left(\frac{d^2}{dR^2} - \frac{l(l+1)}{\hbar^2 R^2} - \hat{W} + P^2 \right) \theta = 0
\label{eq27}
\ee
where $\chi = A(R) \theta$. $A$ is a transition matrix $ A = \exp(-\int^R_{R_0} \tau ^{(1)} (R) dR)$ , where $R_0$ is some large R value where $\phi$ has its asymptotic form. $\hat{W} = \frac{2\mu}{\hbar^2}(\hbar A ^\dag \widetilde{\omega} A)$; P is a 2 $\times$ 2 diagonal matrix whose elements are $P_i =  \sqrt{\frac{2\mu}{\hbar^2}E}$, i=$\alpha, \beta$. Equation (\ref{eq27}) represents scattering equation of two coupled channels in a compact form. For our system, the form of $\hat{W}$-matrix and other related terms are shown in Fig.\ref{fig5}.

\end{document}